\documentclass[sigconf, nonacm]{acmart}

\renewcommand\footnotetextcopyrightpermission[1]{} 
\copyrightyear{2025}

\usepackage{amsmath,amsfonts}
\usepackage{algorithmic}
\usepackage{graphicx}
\usepackage{textcomp}
\usepackage{xcolor}
\usepackage{todonotes}
\usepackage{multirow}
\usepackage{tikz}
\usetikzlibrary{shapes.geometric, arrows, positioning}
\usepackage{pifont}
\usepackage{booktabs}
\usepackage{wrapfig}
\usepackage{url}
\usepackage{array} 

\newcolumntype{C}[1]{>{\centering\arraybackslash}p{#1}}

\begin{document}

\title{UniBOM – A Unified SBOM Analysis and Visualisation Tool for IoT Systems and Beyond}


\author{Vadim Safronov}
\authornote{Equal contribution.}
\email{vadim.safronov@cs.ox.ac.uk}
\affiliation{%
  \institution{University of Oxford}
  \city{Oxford}
  \country{United Kingdom}
}

\author{Ionut Bostan}
\authornotemark[1]
\email{ionut@nquiringminds.com}
\affiliation{%
  \institution{NquiringMinds}
  \city{Southampton}
  \country{United Kingdom}}

\author{Nicholas Allott}
\email{nick@nquiringminds.com}
\affiliation{%
  \institution{NquiringMinds}
  \city{Southampton}
  \country{United Kingdom}
  }

\author{Andrew Martin}
\email{andrew.martin@cs.ox.ac.uk}
\affiliation{%
  \institution{University of Oxford}
  \city{Oxford}
  \country{United Kingdom}
}

\renewcommand{\shortauthors}{Safronov et al.}

\begin{abstract}
Modern networked systems rely on complex software stacks, which often conceal vulnerabilities arising from intricate interdependencies. A Software Bill of Materials (SBOM) is effective for identifying dependencies and mitigating security risks. However, existing SBOM solutions lack precision, particularly in binary analysis and non-package-managed languages like C/C++.

This paper introduces UniBOM, an advanced tool for SBOM generation, analysis, and visualisation, designed to enhance the security accountability of networked systems. UniBOM integrates binary, filesystem, and source code analysis, enabling fine-grained vulnerability detection and risk management. Key features include historical CPE tracking, AI-based vulnerability classification by severity and memory safety, and support for non-package-managed C/C++ dependencies.

UniBOM’s effectiveness is demonstrated through a comparative vulnerability analysis of 258 wireless router firmware binaries and the source code of four popular IoT operating systems, highlighting its superior detection capabilities compared to other widely used SBOM generation and analysis tools. Packaged for open-source distribution, UniBOM offers an end-to-end unified analysis and visualisation solution, advancing SBOM-driven security management for dependable networked systems and broader software.
\end{abstract}



\maketitle

\section{Introduction}

As IoT technologies become increasingly pervasive across diverse safety-critical applications --- ranging from smart cities to industrial IoT and healthcare --- dependable networked systems supporting such infrastructures must navigate increasingly complex software stacks with numerous interdependent components. Ensuring the dependability and security of these systems is a critical challenge, particularly given the distributed nature of IoT applications and the diversity of technological standards they encompass. The high heterogeneity of IoT devices --- spanning variations in types, protocols, manufacturers, update cycles, and software versions --- significantly complicates the task of guaranteeing consistent, high-level protection for IoT applications operating across such diverse environments. These complexities often expose hidden vulnerabilities within interconnected software layers, compromising system dependability, reliability, and security.

Memory-related vulnerabilities in IoT systems account for over 70\% of known security threats~\cite{b3}, a figure consistent with reports from CISA~\cite{CISA} and leading technology organisations~\cite{Microsoft_CVE_report, Chromium_CVE_report, mozilla_rust_rewrite}. When exploited on IoT safety-critical ecosystems, a single compromised router can become the entry point for attacking IoT devices, which subsequently compromises entire networks, escalating to widespread cyber threats. Notable incidents highlight the urgency of this issue: the Dyn DNS attack~\cite{DynDNS} and Mirai botnets~\cite{Mirai-botnet}, where a multitude of IoT devices were used to launch a massive Distributed Denial of Service (DDoS) attacks, the SolarWinds breach~\cite{Solarwinds}, and Log4j incidents~\cite{Log4j} all exemplify the wide systemic consequences of vulnerabilities (however they arise).

Employing Software Bill of Materials (SBOM) advances our ability to develop dependable IoT ecosystems which are both secure and resilient by design. SBOM provides a standardised framework for enumerating software dependencies, enabling tracking of potential vulnerabilities within the analysed software and its subcomponents. Where SBOM concepts are supported by practical tools~\cite{NSA_SBOM, NIST_trusted_IoT_onboarding}, existing and novel threats can, in principle, be the subject of rapid response.  For instance, deploying a robust SBOM-based tool capable of rapidly mapping software dependencies within safety-critical IoT systems affected by a new unknown threat and enabling agile and comprehensive recovery can protect these systems and their interconnected neighbours from the propagation and exploitation of that threat on a global scale.

Despite the availability of various SBOM generation tools~\cite{Trivy_SBOM-tool, Microsoft_SBOM-tool, Syft_SBOM-tool, Intel_sbom-tool}, existing solutions provide limited support for the comprehensive analysis of dependable IoT networked systems where a significant portion of software is written in C/C++, operating without standardised dependency management tools such as Conan~\cite{conan_package_manager}, or relying on frequent binary updates. These limitations arise from the reliance of most SBOM tools on metadata-based approaches, which lack the capability to extract and analyse binary images and C/C++ source code~\cite{b1}. While individual tools dedicated to either binary analysis or source code dependency analysis for C/C++ are available~\cite{b2, b17}, there is no unified end-to-end solution that currently integrates these approaches for generating a fine-grained SBOM for dependable networked systems. Additionally, existing tools often lack advanced visualisation and classification capabilities, such as historical threat analysis, prioritisation of severe vulnerabilities, and comprehensive threat categorisation.

To address these limitations and provide a robust solution for researchers and industry professionals managing the security of modern IoT systems and dependable systems more broadly, we propose UniBOM --- an open source tool for SBOM generation, analysis, and visualisation\footnote{\label{repo}UniBOM CLI Tool: \url{https://github.com/nqminds/UniBOM}}\footnote{\label{repo-vis}UniBOM Web GUI: \url{https://github.com/nqminds/UniBOM-Viewer}}. UniBOM enhances visibility and accountability in dependable system software security by offering the following key functionalities:

\begin{itemize}
    \item \textbf{SBOM generation for binary images:} Supports firmware binary analysis using the built-in Binwalk utility~\cite{b17} to extract firmware filesystems, followed by SBOM generation with Syft~\cite{Syft_SBOM-tool} and CVE/CWE analysis with Grype~\cite{grype}.
    \item \textbf{Tracking non-package-managed C/C++ dependencies:} Facilitates fine-grained SBOM generation by supporting non-conan C/C++ source code analysis, leveraging integration with the CCScanner tool~\cite{b2}.
    \item \textbf{AI-based vulnerability classification:} Addressing the prevalence of memory-safety threats in memory-unsafe languages like C/C++, the tool categorises identified vulnerabilities into memory-related categories and CVSS severity levels, prioritising high-risk vulnerabilities for more fine-grained and effective threat management.
    \item \textbf{Web GUI for visualising advanced security analyses:} A web application that enables users to visualise SBOM analysis results generated by UniBOM or other SBOM tools. It highlights and categorises vulnerabilities based on severity and relevance to memory safety, offering interactive dashboards for vulnerability distribution and historical trend analysis of software components and the emergence of vulnerabilities over time.
\end{itemize}

This paper demonstrates the primary functionalities of UniBOM and is organised as follows: Section 2 provides a concise overview of SBOM concepts. Section 3 discusses related work on SBOM generation tools and methodologies, providing context for UniBOM’s contributions. Section 4 details the tool's design and highlights its main functionalities. Section 5 showcases UniBOM’s capabilities in SBOM generation, emphasising improved vulnerability detection through more precise SBOM creation. This is demonstrated through a case study involving 258 wireless gateway firmware binaries and source code analysis of four widely used IoT operating systems. Section 6 introduces the web application, which presents UniBOM analysis results via interactive dashboards, trend analysis, and risk visualisation, categorising vulnerabilities by severity and memory-safety relevance for comprehensive assessment. Section 7 concludes the paper, summarising key insights and future directions.

\section{Software Bill of Materials}

The Software Bill of Materials (SBOM), inspired by manufacturing industry practices \cite{b10}\cite{b11}, was formalised in 2018 by the National Telecommunications and Information Administration (NTIA) to enhance software security practices, and has since been refined and expanded~\cite{b12}. An SBOM provides a detailed inventory of the components comprising a software application, specifying their origins, dependencies, and associated references to known or potential vulnerabilities.

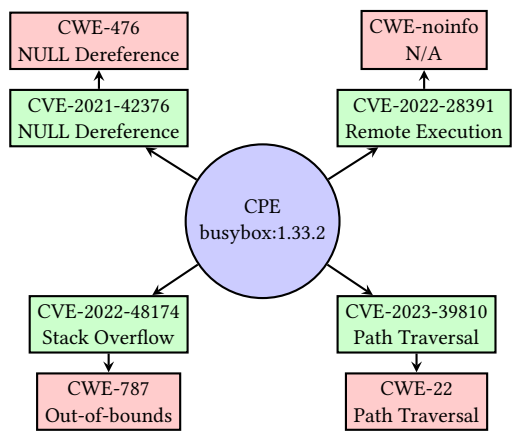
\begin{figure}[ht]
    \centering
    \begin{tikzpicture}[node distance=0.25cm and 0.25cm, >=latex', align=center, font=\small]
        \tikzset{
            central/.style={circle, draw=black, thick, minimum size=0.7em, fill=blue!20},
            cve/.style={rectangle, draw=black, thick, minimum width=1cm, minimum height=0.7em, fill=green!20},
            cwe/.style={rectangle, draw=black, thick, minimum width=1cm, minimum height=0.7em, fill=red!20},
            arrow/.style={thick, ->, >=stealth}
        }

        \node[central] (cpe) {CPE\\busybox:1.33.2};
        
        \node[cve, above left=of cpe] (cve1) {CVE-2021-42376\\NULL Dereference};
        \node[cve, above right=of cpe] (cve2) {CVE-2022-28391\\Remote Execution};
        \node[cve, below left=of cpe] (cve3) {CVE-2022-48174\\Stack Overflow};
        \node[cve, below right=of cpe] (cve4) {CVE-2023-39810\\Path Traversal};
        
        \node[cwe, above=of cve1] (cwe1) {CWE-476\\NULL Dereference};
        \node[cwe, above=of cve2] (cwe2) {CWE-noinfo\\N/A};
        \node[cwe, below=of cve3] (cwe3) {CWE-787\\Out-of-bounds};
        \node[cwe, below=of cve4] (cwe4) {CWE-22\\Path Traversal};

        \draw[arrow] (cpe) -- (cve1);
        \draw[arrow] (cpe) -- (cve2);
        \draw[arrow] (cpe) -- (cve3);
        \draw[arrow] (cpe) -- (cve4);
        
        \draw[arrow] (cve1) -- (cwe1);
        \draw[arrow] (cve2) -- (cwe2);
        \draw[arrow] (cve3) -- (cwe3);
        \draw[arrow] (cve4) -- (cwe4);
    \end{tikzpicture}
    \caption{Radial representation of CPE, CVE, and CWE relationships for Busybox 1.33.2.}
    \label{fig:busybox-cpe-cve-cwe}
\end{figure}

Key foundational elements within the SBOM ecosystem include the Common Platform Enumeration (CPE), Common Vulnerabilities and Exposures (CVE), and Common Weakness Enumeration (CWE). Figure~\ref{fig:busybox-cpe-cve-cwe} illustrates the connections between these components, facilitating the identification, classification, and mitigation of security flaws and vulnerabilities. An SBOM, essentially a list of software components (CPEs), ensures that all elements are identified and traced to their origins.

\textbf{CPE} is a standardised naming schema maintained by the U.S. National Institute of Standards and Technology (NIST)~\cite{b5}, designed to identify software versions and packages. The current version, CPE 2.3, assigns structured names to software products, capturing attributes such as part, vendor, product, and version\cite{b6}. For example, the GNOME GLib library version 2.66.4 is encoded as \verb|cpe:2.3:a:gnome:glib:2.66.4:::::::*|, where ``cpe:2.3" indicates the CPE version (2.3), ``a" specifies the application component, ``gnome" identifies the vendor, ``glib" is the product, and ``2.66.4" denotes the version.

\textbf{CVE} provides a unified system to identify, assess, and catalogue vulnerabilities. Hosted on the National Vulnerability Database (NVD)~\cite{b5}, it assigns unique IDs to facilitate vulnerability tracking, discussion, and resolution~\cite{b7}. For instance, CVE-2021-42376, a NULL Dereference vulnerability, relates to the product identified as \verb|cpe:2.3:a:busybox:busybox:1.33.2|. A CVE typically categorises security vulnerabilities in software through key components. The CVE ID (``CVE-YYYY-NNNNN") includes ``YYYY," representing the year of discovery, and ``NNNNN," a unique identifier for the vulnerability. The description provides details about the vulnerability and its potential impact on affected systems. Weaknesses identify the type of security issue (CWE). The base score quantifies the severity of the vulnerability, while the base severity provides a qualitative rating as ``Low," ``Medium," ``High," or ``Critical".

\textbf{CWE} is a comprehensive glossary of common security weaknesses in software and hardware, with updates made several times a year \cite{b9}. Each CWE may encompass multiple CVE instances, as it categorises a specific type of weakness that can lead to various vulnerabilities. For example, CWE-787, Out-of-Bounds Write, has been the root cause of numerous vulnerabilities, such as the stack overflow vulnerability in BusyBox version 1.33.2 (CVE-2021-42373) and a memory corruption issue in OpenSSL (CVE-2022-0778). 

Overall, CPEs identify software elements across systems, CVEs link these elements to known vulnerabilities, and CWEs provide insight into potential weak points within internal components. 

\section{Related Work}
SBOM generation and analysis methods can be broadly categorised into binary analysis, metadata-based analysis, and source code inspection.

Binary-focused SBOM generation tools~\cite{B2SFinder, b17, OSSPolice, BAT} perform software composition analysis on compiled binaries, leveraging string literals, embedded metadata, and language-specific features to identify software components and their dependencies.

Metadata-based SBOM generation tools~\cite{Trivy_SBOM-tool, Microsoft_SBOM-tool, Syft_SBOM-tool, cdxgen} analyse software package metadata and dependency information. These tools support diverse ecosystems by parsing metadata from container images, build files, and package managers to map dependencies effectively.

Source code analysis tools~\cite{b14, tamer, centris} generate SBOMs by examining code repositories to identify components, dependencies, and their interrelationships. Compared to binary or metadata-focused approaches, they can provide more detailed SBOM insights by uncovering hidden interdependencies, including those potentially associated with present CVEs.

UniBOM distinguishes itself by integrating all three methods: binary extraction via Binwalk~\cite{b17}, filesystem and metadata analysis through Syft~\cite{Syft_SBOM-tool}, and source code analysis using CCScanner~\cite{b14}. Additionally, UniBOM's web-based GUI visualises SBOM analysis, highlighting vulnerabilities by severity and memory safety relevance, with interactive dashboards for distribution and historical trends of components and vulnerabilities. 

While a few visualisation tools for SBOM analysis are available~\cite{fossa_sbom,deepbits_tools}, their analytical capabilities are limited in scope and do not include features such as historical vulnerability analysis or vulnerability categorisation based on memory-safety relevance, which UniBOM aims to provide as part of a more comprehensive approach.

\section{UniBOM Architecture}
\label{design}

\begin{figure*}[t]
\centering
    \includegraphics[scale=0.35]{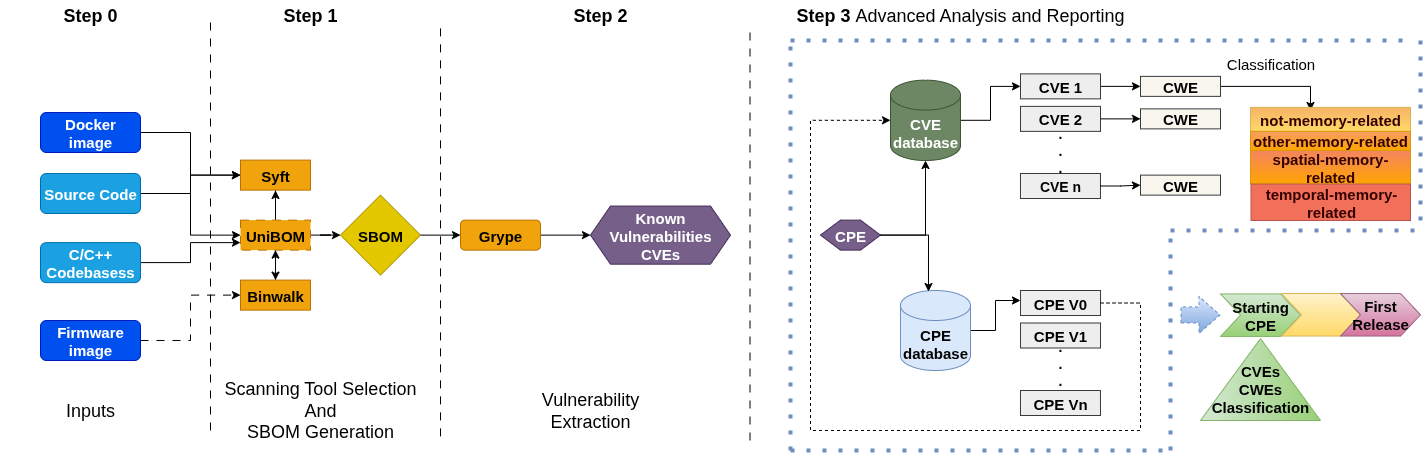}
\caption{UniBOM architectural pipeline.}
\label{fig:architecture}
\end{figure*}

UniBOM is a Node.js-based command-line interface (CLI) tool designed for comprehensive SBOM generation and vulnerability analysis across a wide range of input types, including Docker images, source code, root filesystems, C/C++ codebases, and firmware binaries. Its architecture integrates various tools for filesystem extraction, code and filesystem scanning, SBOM generation and analysis, as well as advanced CVE analysis and categorisation, including Binwalk~\cite{b17}, Syft~\cite{Syft_SBOM-tool}, Grype~\cite{grype}, and CCScanner~\cite{b14},  all packaged in Docker containers for simplified installation and usability. By combining these capabilities into an end-to-end architectural pipeline~(Figure~\ref{fig:architecture}), UniBOM offers a professional-grade solution for SBOM generation and vulnerability analysis, supporting both research and industrial security applications and effectively addressing complex and diverse input types with precision. The UniBOM pipeline operates through the following steps:

\begin{itemize}
    \item \textbf{Step 0: Input Format Identification.} The pipeline begins by recognising the input format, which can range from container images to binary firmware or source code.
    \item \textbf{Step 1: Tool Selection and SBOM generation.} Based on the input format and desired SBOM granularity, UniBOM selects the appropriate processing tool for SBOM generation, leveraging Syft for general-purpose SBOM and CCScanner for C/C++ codebases, addressing limitations in recognising third-party dependencies without a package manager.
    \item \textbf{Step 2: CVE/CWE Extraction.} For each dependency in the generated SBOM, UniBOM performs vulnerability detection by mapping identified components and dependencies to the CPE v2.3 format and, through requesting vulnerability databases, extracting of associated CVEs and their general CWEs, for overall report and further advanced security analysis.
    \item \textbf{Step 3: Advanced Security Analysis.} Beyond SBOM generation and CVE/CWE detection, UniBOM provides advanced capabilities, including:
        \begin{itemize}
            \item Historical tracing of vulnerabilities for identified CPEs.
            \item Memory safety assessments for detected CVEs and CWEs, offering deeper insights into potential security impacts.
            \item Cross-version SBOM comparisons to track changes and identify evolving risks.
        \end{itemize}
\end{itemize}

\subsection{Usage Overview}

\subsubsection{Binary Extraction}
The CLI tool incorporates Binwalk, a specialised utility for reverse engineering and extracting data from firmware binary images. 

\textbf{\texttt{unibom -binwalk "\$(pwd)" "[-Me]" image.bin}}. 
Binwalk efficiently identifies and unpacks embedded file systems, libraries, executables, and other components, including recognising compression and encryption methods to uncover hidden or obfuscated data. It also extracts metadata, such as OS versions and build environments, facilitating detailed dependency mapping and source code analysis. These capabilities enable the generation of fine-grained SBOMs and advanced security assessments, such as identifying firmware-specific vulnerabilities and CWEs.

\subsubsection{SBOM Generation}
SBOM generation in UniBOM is supported through two distinct methods, tailored to the characteristics of the input data:

\textbf{\texttt{unibom -generateSbom <project\_path> <project\_name>}}
  The integrated Anchore/Syft generator creates an SBOM by analysing extracted filesystems and source code, which can be retrieved directly from a repository or extracted using Binwalk. Syft identifies dependencies managed by supported package managers, such as package.json for Node.js or Conan for C/C++. A detailed list of supported package managers and programming languages is available in Syft’s official documentation~\cite{Syft_SBOM-tool}. To assess vulnerabilities, the SBOM is analysed using Grype, which produces a software vulnerability report based on the generated SBOM.

\textbf{\texttt{unibom -generateCCPPReport <path\_to\_c/cpp\_project>\\<project\_name>}}
This method addresses the limitations of existing tools in generating SBOMs for C/C++ projects lacking standard package management files, such as *.vcxproj or conan.lock. Leveraging insights from large-scale differential analyses of SBOM generators --- including Syft, Trivy, GitHub, and sbom-tool --- UniBOM integrates CCScanner, a research-driven tool designed to unify and standardise dependency analysis for C/C++ environments. CCScanner excels in parsing complex build systems, such as Makefiles, CMake, and Bazel, and extracting dependency information directly from source code, linking headers, and linked libraries. It maps these dependencies to standardised identifiers, such as Common Platform Enumeration (CPE) entries, and facilitates the inclusion of third-party libraries often overlooked due to disjointed package management in C/C++ projects. This integration enables UniBOM to produce more fine-grained SBOMs that account for both direct and transitive dependencies, enhancing the visibility of software components and improving vulnerability assessment in firmware with the majority of C/C++ sourcecode.

\subsubsection{Extended Analysis and Comparison Features} 

UniBOM's advanced SBOM analysis features offer detailed insights into the history of CPE releases and their associated vulnerability analysis, categorising security weaknesses for enhanced clarity. These features also enable direct SBOM comparisons to track changes and progress over time, such as comparing neighbouring firmware updates. The following advanced analysis functionality is designed to provide UniBOM users with a comprehensive understanding of the security state of their firmware, enabling them to make well-informed decisions on further targeted improvements and their prioritisation.

\textbf{\texttt{unibom -getHistory <CPE>}}
This parameter enables the historical analysis of CPE components. For a specified CPE, the tool retrieves and examines all prior versions of that CPE. For each version, it displays relevant information, such as associated CVEs and, for each CVE, its corresponding CWE. It further categorises each CWE based on memory-related classifications (e.g., spatial, temporal, or other memory types). The structure, as illustrated in Table~\ref{tab:cpe_vulnerabilities}, provides a detailed security taxonomy for every version, enabling users to identify the vulnerabilities present in their code and understand their impact on memory usage. This functionality is particularly valuable in environments coded in C/C++ languages, where memory management is a critical consideration. The ability to monitor such intricate interconnections over time is advantageous for proactive risk management. It allows developers and security teams to adaptively assess the risk levels of software components and prioritise mitigation efforts accordingly. This comprehensive perspective facilitates more informed decisions about securing dependencies, especially when supporting legacy versions or managing multiple library versions with varying levels of known vulnerabilities.

\begin{table}[htbp]
\caption{CPE historical analysis.}
\begin{center}
\begin{tabular}{|c|c|c|p{1.3cm}|}
\hline
\textbf{CPE} & \textbf{CVE} & \textbf{CWE} & \textbf{Memory class} \\ \hline
\multicolumn{4}{|c|}{\ldots (multiple versions)} \\ \hline
\shortstack{cpe:2.3:a:openssl \\ openssl:0.9.2b} 
  & CVE-2014-8176 & CWE-119 & spatial \\ \hline
\multicolumn{4}{|c|}{\ldots (multiple versions)} \\ \hline
\shortstack{cpe:2.3:a:openssl \\ openssl:0.9.6d} 
  & CVE-2016-2106 & CWE-189 & spatial \\ \hline
\multicolumn{4}{|c|}{\ldots (multiple versions)} \\ \hline
\multirow{3}{*}{\shortstack{cpe:2.3:a:openssl \\ openssl:1.1.1}} 
  & CVE-2021-3712 & CWE-125 & spatial \\ \cline{2-4}
  & CVE-2022-4450 & CWE-415 & temporal \\ \cline{2-4}
  & CVE-2021-3449 & CWE-476 & spatial \\ \hline
\end{tabular}
\label{tab:cpe_vulnerabilities}
\end{center}
\end{table}

\textbf{\texttt{unibom -classifyCwe <CWE-ID>}}
UniBOM leverages a GPT-based model to classify vulnerabilities into four categories: not memory-related, spatial memory-related, temporal memory-related, and other memory-related. This is achieved by sending a structured prompt with instructions for classifying the vulnerability based on its textual description. If the CWE text description is unavailable, the corresponding CVE description can be used for classification. Understanding and quantifying memory-related vulnerabilities is essential for making informed decisions, such as adopting hardware-based memory-protection solutions like the CHERI-based Morello architecture~\cite{b4} or transitioning to a memory-safe programming language like Rust. The prevalence of memory vulnerabilities in IoT devices, as highlighted in a recent study on IoT firmware analysis~\cite{b3}, emphasises the importance of classifying these risks. IoT devices are particularly susceptible to a range of memory-safety threats, including spatial vulnerabilities such as buffer overflows and temporal vulnerabilities like use-after-free errors. Without quantifying and categorising these vulnerabilities, it is difficult to assess their full impact on system security. By identifying and classifying these risks, stakeholders can gain a clearer understanding of the scale and severity of memory-related issues within the analysed firmware, enabling more informed and effective decision-making.

\textbf{\texttt{unibom -compare <sbom1> <sbom2>}} The SBOM compare function provides an in-depth analysis of multiple SBOMs by comparing component versions and their associated CVEs. This feature provides fine-grained visibility into the security state across configurations and environments by isolating component version changes and mapping vulnerabilities to specific versions within each SBOM. It is particularly valuable for vulnerability management and compliance reporting, as it offers a clear understanding of changes in dependencies and their impact on overall security. Automating the SBOM comparison process reduces the manual effort required to identify systems or environments at greater risk, while also highlighting configuration-specific gaps or inconsistencies. This enables security teams to make prompt, informed decisions on prioritising remediation efforts.

\begin{table}[h!]
\centering
\caption{SBOM comparison for key CPEs.}
\begin{tabular}{|l|c|c|}
\hline
\textbf{Component}           & \textbf{SBOM-1}   & \textbf{SBOM-2}   \\ \hline

\multicolumn{3}{|c|}{\textbf{Version Information}} \\ \hline
\textbf{openssl}             & 3.0.0             & none           \\ \hline
\textbf{kernel}              & 2.24.2            & 2.24.2            \\ \hline
\textbf{sqlite}              & none           & 3.5.9             \\ \hline
\multicolumn{3}{|c|}{\textbf{Vulnerabilities}} \\ \hline
\textbf{openssl}             &  CVE-2009-1390, ...             & none           \\ \hline
\textbf{kernel}              & CVE-2014-9114, ...             & CVE-2016-2779, ...            \\ \hline
\textbf{sqlite}              & none           & CVE-2015-3414, ...             \\ \hline
\end{tabular}
\label{tab:sbom_comparisson}
\end{table}

Table~\ref{tab:sbom_comparisson} presents an example of the output generated from comparing components identified in two test SBOMs of the same firmware. The CPEs are categorised into ``Component", ``Version Information", and ``Vulnerabilities", highlighting differences between the two SBOMs in terms of component presence, version consistency, and potential security risks.

\begin{itemize}
    \item \textit{Key Components}: indicates the presence or absence of specific components, such as OpenSSL, kernel, and SQLite, in each SBOM. For example, SBOM-1 contains OpenSSL, whereas SBOM-2 does not. Conversely, SQLite is unique to SBOM-2, while the kernel is present in both SBOMs.
    \item \textit{Version Information}: details the components included in each SBOM along with their respective versions. For instance, SBOM-1 lists OpenSSL as version 3.0.0, but SBOM-2 does not include it. The kernel is version 2.24.2 in both SBOMs, while SBOM-2 lists SQLite as version 3.5.9.
    \item \textit{Vulnerabilities}: lists the known vulnerabilities associated with each SBOM component. For example, OpenSSL in SBOM-1 has the vulnerability CVE-2009-1390, but OpenSSL is absent from SBOM-2, and therefore no vulnerabilities are listed for it. The kernel in SBOM-1 has CVE-2014-9114, while SBOM-2 contains CVE-2016-2779 for the same component. Additionally, SQLite vulnerabilities unique to SBOM-2 include CVE-2015-3414.
\end{itemize}

\section{Evaluation}
In order to demonstrate UniBOM's effectiveness in handling the most commonly used input formats of IoT system firmware  --- binary images, filesystems, and source code --- its functionality was compared against popular SBOM tools, including Microsoft's SBOM-tool~(referred to hereafter as sbom-tool)~\cite{Microsoft_SBOM-tool}, Trivy~\cite{Trivy_SBOM-tool}, Syft~\cite{Syft_SBOM-tool}, and CCScanner~\cite{b14}. Two experiments were conducted: (a) SBOM generation followed by vulnerability scanning of real IoT firmware binaries from a recent dataset used in prior CVE analyses of IoT firmware images~\cite{b16}, and (b) source code analysis of popular IoT operating systems.

\subsection{IoT Binary Firmware Analysis}
The evaluation involved 258 Linux-based firmware binaries from leading wireless gateway manufacturers: 54 from D-Link, 104 from OpenWrt, and 100 from TRENDnet.

To ensure a fair comparison with other tools, all binary images were extracted prior to analysis. The extraction was performed using Binwalk~\cite{b17}, and for each extracted firmware binary, SBOMs were generated using Syft, Trivy, and sbom-tool. These SBOMs were subsequently parsed with the Grype vulnerability scanner~\cite{grype} to identify CVEs and analyse their severity occurrences.

UniBOM was the only tool that streamlined the entire process of SBOM generation and vulnerability analysis by integrating Binwalk, Syft, CCScanner, and Grype into a unified workflow, providing seamless binary analysis and vulnerability scanning.

\begin{table}[h!]
\centering
\setlength{\tabcolsep}{2pt} 
\setlength{\abovecaptionskip}{5pt}
\setlength{\belowcaptionskip}{5pt}
\caption{Vulnerability scanning results for IoT firmware binaries.}
\label{tab:vulnerability_images}
\begin{tabular}{|p{1.5cm}|c|c|c|c|C{1.35cm}|}
\hline
\textbf{Tools}       & \textbf{D-Link} & \textbf{OpenWrt} & \textbf{TRENDnet} & \textbf{Total} & \textbf{\begin{tabular}[c]{@{}c@{}}Total\\ Severities\end{tabular}} \\ \hline
\textbf{sbom-tool}   & 0               & 0                & 0                 & 0              & -                                                                   \\ \hline
\textbf{Trivy}       & 0               & 0                & 0                 & 0              & -                                                                   \\ \hline
\textbf{Syft}        & 1885            & 1371             & 1326              & 4582           & \begin{tabular}[c]{@{}l@{}}
526 Crit \\ 
2300 High \\ 
1642 Med \\ 
114 Low 
\end{tabular}                                                           \\ \hline
\textbf{CCScanner}   & 0               & 0                & 0                 & 0              & -                                                                   \\ \hline
\textbf{UniBOM}      & 1885            & 1371             & 1326              & 4582           & \begin{tabular}[c]{@{}l@{}}
526 Crit \\ 
2300 High \\ 
1642 Med \\ 
114 Low 
\end{tabular}                                                           \\ \hline
\end{tabular}
\end{table}

Table \ref{tab:vulnerability_images} shows that UniBOM and Syft were the only tools capable of successfully analysing the extracted filesystems, whereas other tools failed to identify the majority of dependencies within the extracted filesystem data. Both UniBOM and Syft detected a total of 4582 CVE occurrences, with over 60\% classified as high or critical severity. This highlights the advantage of UniBOM’s seamless integration with Binwalk, Syft, CCScanner, and Grype, where Binwalk handles binary extraction, Syft effectively analyses the resulting filesystems (with CCScanner for source code analysis, if required), and Grype performs further vulnerability assessment. 

\subsection{IoT OS Source Code Analysis}

To demonstrate UniBOM’s capabilities beyond binary analysis, its SBOM generation and analysis functionalities were tested on source code from widely used IoT operating systems, including the Raspberry Pi Linux Kernel (RPi)~\cite{raspberrypi_linux}, Zephyr~\cite{zephyr_repo}, Nuttx~\cite{nuttx_repo}, OpenWrt 23.05.5~\cite{openwrt_23055}, and the latest version of OpenWrt~\cite{openwrt_latest}. The evaluation focused on assessing UniBOM’s effectiveness in identifying vulnerabilities through source code analysis. Table~\ref{tab:iot_vulnerabilities} summarises the vulnerabilities detected in the IoT OS source code by the evaluated SBOM analysis tools.

\begin{table}[h!]
\centering
\setlength{\tabcolsep}{2pt} 
\setlength{\abovecaptionskip}{5pt}
\setlength{\belowcaptionskip}{5pt}
\caption{Vulnerability scanning results for IoT OS source code.}
\label{tab:iot_vulnerabilities}
\begin{tabular}{|p{1.5cm}|C{0.48cm}|C{0.97cm}|C{0.77cm}|C{1.29cm}|C{1.29cm}|C{0.97cm}|}
\hline
\textbf{Tools}       & \textbf{RPi} & \textbf{Zephyr} & \textbf{Nuttx} & \textbf{\begin{tabular}[c]{@{}c@{}}OpenWrt\\ 23.05\end{tabular}} & \textbf{\begin{tabular}[c]{@{}c@{}}OpenWrt\\ Latest\end{tabular}} & \textbf{\begin{tabular}[c]{@{}c@{}}Total\\Sevs\end{tabular}} \\ \hline
\textbf{sbom-tool}   & 0            & 0               & 0              & 0                                                                  & 0                                                                  & -                                                                    \\ \hline
\textbf{Trivy}       & 0            & 5               & 0              & 0                                                                  & 0                                                                  & \begin{tabular}[c]{@{}l@{}}
2 Crit \\ 
2 Med \\ 
1 Low
\end{tabular}                                                          \\ \hline
\textbf{Syft}        & 0            & 5               & 0              & 0                                                                  & 4                                                                  & \begin{tabular}[c]{@{}l@{}}
2 Crit \\ 
2 High \\ 
5 Med
\end{tabular}                                                          \\ \hline
\textbf{CCScanner}   & 2            & 0               & 38             & 1                                                                  & 0                                                                  & \begin{tabular}[c]{@{}l@{}}
9 Crit \\ 
15 High \\ 
21 Med \\ 
1 Low \\ 
1 Unkn
\end{tabular}                                                          \\ \hline
\textbf{UniBOM}      & 2            & 5               & 38             & 1                                                                  & 4                                                                  & \begin{tabular}[c]{@{}l@{}}
11 Crit \\ 
17 High \\ 
26 Med \\ 
1 Low \\ 
1 Unkn
\end{tabular}                                                          \\ \hline
\end{tabular}
\end{table}

UniBOM significantly outperformed all other standalone tools by integrating Syft’s metadata extraction with CCScanner’s advanced source code analysis techniques. This combination allowed UniBOM to detect a broader range of dependencies and uncover more vulnerabilities, identifying the highest number of CVE occurrences, with a total of 56. Compared to CCScanner, UniBOM identified 2 additional critical, 5 medium, and 2 high-severity CVEs. Against Syft, UniBOM detected 2 more critical, 15 high-severity, and 21 medium-severity CVEs. The performance gap was even more evident when compared to Trivy and SBOM-tool, which detected only 5 and 0 CVEs, respectively.

\section{Web GUI Overview}

The UniBOM GUI, powered by its SBOM generation and analysis engine as detailed in Section~\ref{design}, is designed to provide convenience in analysis visualisation, enhancing user understanding of the security state and trends of analysed firmware.
\begin{figure*}[t]
\centering
\includegraphics[scale=0.23]{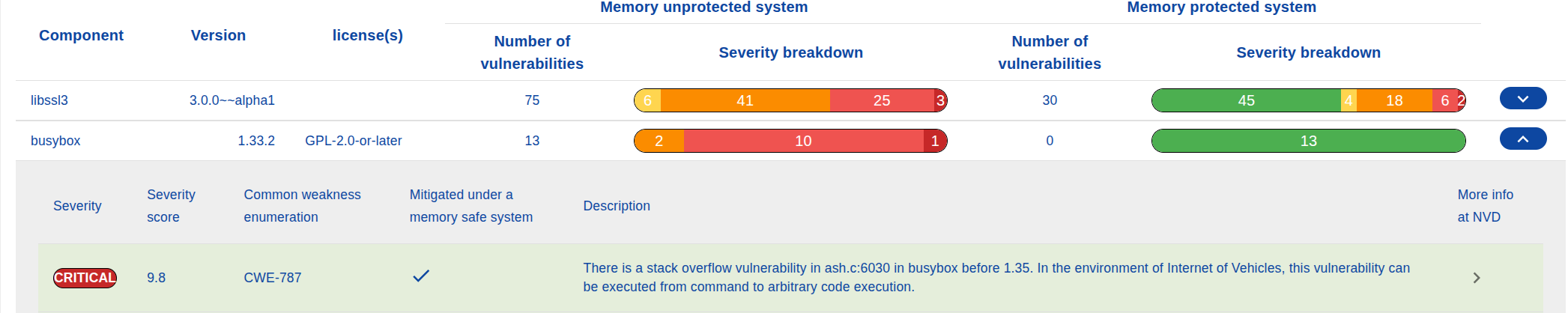}
\caption{UniBOM GUI screenshot: firmware vulnerability status.}
\label{fig:gui_status}
\end{figure*}
\begin{figure*}[htbp]
\centering
\includegraphics[width=\textwidth]{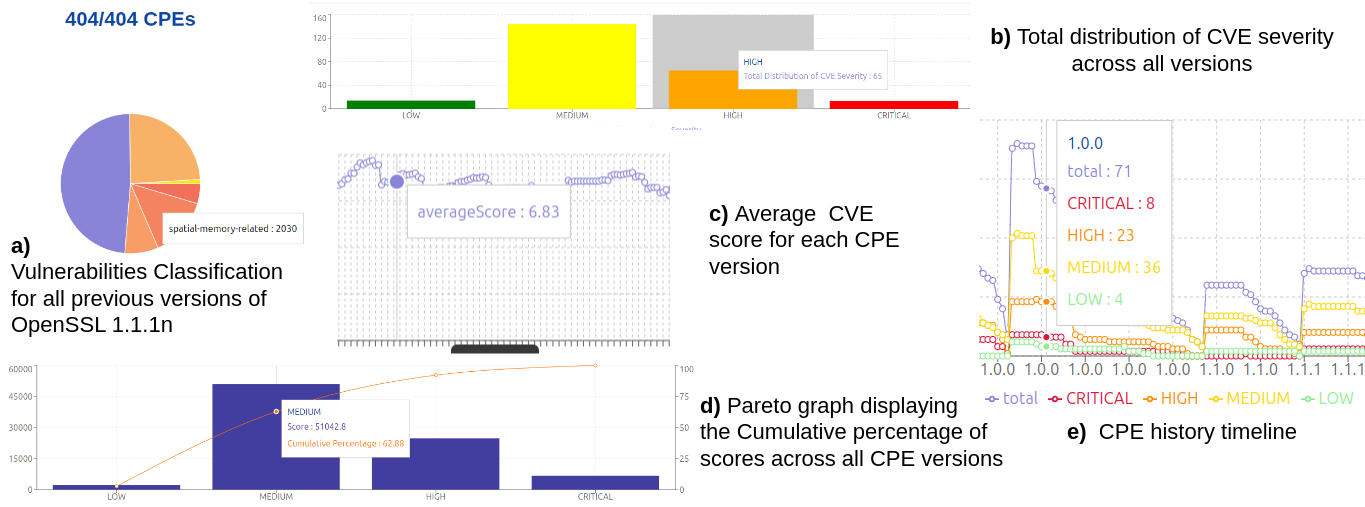}
\caption{UniBOM GUI screenshot: historical analysis for OpenSSL 1.1.1n.}
\label{fig:second_view}
\end{figure*}
Users can upload SBOM files for in-depth analysis, with results presented through interactive dashboards and charts. Key features include listing all components within the SBOM file, identifying vulnerabilities associated with each component, and categorising these vulnerabilities based on severity and relevance to memory-safety issues, which are prevalent in IoT systems. To make multidimensional analysis data accessible, the UniBOM GUI employs dynamic bar charts to illustrate the distribution of vulnerabilities by severity, pie charts to classify vulnerabilities based on memory safety, and tables to offer detailed information on each vulnerability, including severity, affected systems, and links to external National Vulnerability Database (NVD) references. The GUI includes a historical analysis feature, offering insights into CVE appearance and mitigation trends across selected CPEs and their previous versions.

\subsection{Visualising Firmware Vulnerability Status}
When an SBOM file is uploaded, UniBOM enumerates all its CPEs and groups the associated CVEs by severity level (Low, Medium, High, or Critical) and memory-safety classification (memory-related or non-memory-related). These classifications are visualised through pie charts, as shown in Figure~\ref{fig:gui_status}, providing users with an overview of the risk distribution within the analysed firmware.

Figure~\ref{fig:gui_status} also highlights memory classification and an analysis of the potential impact of transitioning selected firmware to secure-by-design memory protection solutions, such as CHERI~\cite{CHERI}, CHERIoT~\cite{CHERIoT}, Rust~\cite{Rust} and other potential memory-protection approaches discussed in recent works~\cite{tagged_memory_protection_survey, infat_pointer, spectre_era_mitigations}. The screenshot in Figure~\ref{fig:gui_status} showcases an analysis of example firmware, demonstrating that adopting a memory protection solution could eliminate 42 medium, high, and critical CVE occurrences, significantly reducing the vulnerability surface. This feature is particularly critical for IoT systems, where memory-safety threats are notably widespread.

\subsection{Historical Analysis: Understanding Vulnerability Trends}

UniBOM's historical analysis capability is one of its most distinctive features, enabling users to examine the vulnerability trajectory of specific components over time (Figure~\ref{fig:second_view}). A user can select a specific CPE, e.g. OpenSSL 1.1.1n, and retrieve all previously identified CPE versions of that component along with their associated vulnerabilities (Figure~\ref{fig:second_view}b, c, and e).

The severity distribution graphic uses intuitive colour coding to help users quickly identify components with the highest risk levels. Similarly, the memory vulnerability breakdown chart highlights whether the CPE weaknesses are due to spatial or temporal memory issues (Figure~\ref{fig:second_view}a).

Time-series charts display the number of vulnerabilities for each component version over time. For instance, the OpenSSL analysis highlights a spike in vulnerabilities for specific versions (Figure~\ref{fig:second_view}e), enabling users to identify when and where security issues were introduced. Additionally, Figure~\ref{fig:second_view}c shows the average severity score for each previous OpenSSL versions, allowing for deeper insights into version-specific risks.

Pareto charts aggregate vulnerabilities by severity, helping users focus on the most critical issues that account for the majority of vulnerabilities (Figure~\ref{fig:second_view}d).

The historical analysis feature not only provides insights into past security trends but, through extrapolation, can also help anticipate and predict potential vulnerabilities in future updates.

\section{Conclusion}

This paper introduced UniBOM, a comprehensive tool for SBOM generation, analysis, and visualisation, designed to address the challenges of ensuring the security and accountability of IoT systems and beyond. Its web application enhances usability with interactive dashboards, historical trend analysis, and detailed risk visualisation.

UniBOM addresses the limitations of existing SBOM tools by integrating binary, filesystem, and source code analysis into a unified framework, enabling a more fine-grained approach to SBOM creation and seamless vulnerability analysis. Its ability to categorise vulnerabilities by severity and memory-safety relevance enables the effective prioritisation of protective actions, particularly in systems utilising memory-unsafe languages such as C/C++, which remain prevalent in critical systems. A case study involving 258 wireless gateway firmware binary images and the source code of four popular IoT operating systems demonstrated UniBOM’s superior detection capabilities through more precise SBOM generation compared to existing solutions.

Although the demonstration examples focus on IoT networked systems security, UniBOM is a general-purpose SBOM tool, and thus applies to diverse software environments. Future work will prioritise embedding UniBOM into operational workflows and introducing further enhancements to improve its effectiveness and strengthen its impact on the security accountability of real-world software systems. In this direction, we plan to integrate UniBOM with AI Bill of Materials (AIBOM) approaches, such as TAIBOM~\cite{taibom}, to strengthen security, accountability, and trust in both modern and next-generation AI-enabled software systems.

\begin{acks}
This work was supported by the Innovate UK-funded Secure Networking by Design (SNbD) project, grant number 10028034.
\end{acks}

\bibliographystyle{ACM-Reference-Format}
\bibliography{references}

@INPROCEEDINGS{b1,
  author={Yu, Sheng and Song, Wei and Hu, Xunchao and Yin, Heng},
  booktitle={2024 54th Annual IEEE/IFIP International Conference on Dependable Systems and Networks (DSN)}, 
  title={On the Correctness of Metadata-Based SBOM Generation: A Differential Analysis Approach}, 
  year={2024},
  volume={},
  number={},
  pages={29-36},
  doi={10.1109/DSN58291.2024.00018}}

@inproceedings{b2,
  author = {W. Tang and Z. Xu and C. Liu and J. Wu and S. Yang and Y. Li and P. Luo and Y. Liu},
  title = {Towards Understanding Third-party Library Dependency in C/C++ Ecosystem},
  booktitle = {Proceedings of the 37th IEEE/ACM International Conference on Automated Software Engineering (ASE ’22)},
  year = {2022},
  month = {Oct.},
  pages = {1--12},
  url = {http://dx.doi.org/10.1145/3551349.3560432}
}

@inproceedings{b3,
author = {Safronov, Vadim and Bostan, Ionut and Allott, Nicholas and Martin, Andrew},
title = {How Memory-Safe is IoT? Assessing the Impact of Memory-Protection Solutions for Securing Wireless Gateways},
year = {2025},
isbn = {9798400712852},
publisher = {Association for Computing Machinery},
address = {New York, NY, USA},
url = {https://doi.org/10.1145/3703790.3703820},
doi = {10.1145/3703790.3703820},
booktitle = {Proceedings of the 14th International Conference on the Internet of Things},
pages = {261–266},
numpages = {6},
location = {
},
series = {IoT '24}
}

@article{b4,
  author = {R. Grisenthwaite and G. Barnes and R. N. M. Watson and S. W. Moore and P. Sewell and J. Woodruff},
  title = {The Arm Morello Evaluation Platform—Validating CHERI-Based Security in a High-Performance System},
  journal = {IEEE Micro},
  volume = {43},
  number = {3},
  pages = {50--57},
  year = {2023},
  url = {http://dx.doi.org/10.1109/MM.2023.3264676}
}

@misc{b5,
  author = {NIST CPE},
  title = {{NVD - Common Platform Enumeration (CPE)}},
  year = {2024},
  url = {https://nvd.nist.gov/products/cpe},
  note = {[Accessed: Jun. 19, 2024]}
}

@misc{b6,
  author = {MITRE},
  title = {{CPE Version 2.3 Specifications}},
  year = {2024},
  url = {https://cpe.mitre.org/specification/},
  note = {{[Accessed: Jun. 19, 2024]}}
}

@misc{b7,
  author = {NIST CVE},
  title = {NIST's CVE Process},
  year = {2024},
  url = {https://nvd.nist.gov/general/cve-process},
  note = {[Accessed: Jun. 19, 2024]}
}

@misc{b9,
  author = {MITRE Corporation},
  title = {About Common Weakness Enumeration},
  year = {2024},
  url = {https://cwe.mitre.org/about/index.html},
  note = {[Accessed: Jun. 19, 2024]}
}

@article{b10,
  author = {B. Xia and T. Bi and Z. Xing and Q. Lu and L. Zhu},
  title = {An Empirical Study on Software Bill of Materials: Where We Stand and the Road Ahead},
  journal = {arXiv},
  year = {2023},
  url = {https://arxiv.org/abs/2301.05362}
}

@article{b11,
  author = {R. Jiao and M. Tseng and Q. Ma and Y. Zou},
  title = {Generic Bill-of-Materials-and-Operations for High-Variety Production Management},
  journal = {Concurrent Engineering: Research and Applications},
  volume = {8},
  number = {4},
  pages = {297--321},
  month = {Dec.},
  year = {2000},
  url = {https://doi.org/10.1177/1063293X0000800404}
}

@misc{b12,
  author = {J. Biden},
  title = {Executive Order on Improving the Nation's Cybersecurity},
  month = {May},
  year = {2021},
  url = {https://www.whitehouse.gov/briefing-room/presidential-actions/2021/05/12/executive-order-on-improving-the-nations-cybersecurity/},
  note = {[Accessed: Jun. 20, 2024]}
}

@article{b14,
  author = {W. Tang and Z. Xu and C. Liu and J. Wu and S. Yang and Y. Li and P. Luo and Y. Liu},
  title = {Towards Understanding Third-party Library Dependency in C/C++ Ecosystem},
  booktitle = {Proceedings of the 37th IEEE/ACM International Conference on Automated Software Engineering (ASE)},
  month = {Oct.},
  year = {2022},
  url = {https://doi.org/10.1145/3551349.3560432}
}

@inproceedings{b16,
  author = {B. Zhao and S. Ji and J. Xu and Y. Tian and Q. Wei and Q. Wang and C. Lyu and X. Zhang and C. Lin and J. Wu and R. Beyah},
  title = {A Large-Scale Empirical Analysis of the Vulnerabilities Introduced by Third-Party Components in IoT Firmware},
  booktitle = {Proceedings of the 31st ACM SIGSOFT International Symposium on Software Testing and Analysis (ISSTA)},
  month = {Jul.},
  year = {2022},
  pages = {442--454},
  url = {https://doi.org/10.1145/3533767.3534366}
}

@misc{b17,
  author = {ReFirm Labs},
  title = {Binwalk},
  year = {2024},
  url = {https://github.com/ReFirmLabs/binwalk},
  note = {[Accessed: Aug. 5, 2024]}
}

@misc{Microsoft_CVE_report,
  author = {Microsoft Security Response Center (MSRC)},
  title = {A Proactive Approach to More Secure Code},
  year = {2019},
  month = {Jul.},
  url = {https://msrc.microsoft.com/blog/2019/07/a-proactive-approach-to-more-secure-code/},
  note = {[Accessed: Nov. 28, 2024]}
}

@misc{Chromium_CVE_report,
  author = {The Chromium Project},
  title = {Memory Safety - Chromium Security},
  year = {n.d.},
  url = {https://www.chromium.org/Home/chromium-security/memory-safety/},
  note = {[Accessed: Nov. 28, 2024]}
}

@misc{mozilla_rust_rewrite,
  author = {Mozilla Hacks},
  title = {Rewriting a Browser Component in Rust},
  year = {2019},
  month = {Feb.},
  url = {https://hacks.mozilla.org/2019/02/rewriting-a-browser-component-in-rust/},
  note = {[Accessed: Nov. 28, 2024]}
}

@misc{CISA,
  author = {Cybersecurity and Infrastructure Security Agency (CISA)},
  title = {The Urgent Need for Memory Safety in Software Products},
  year = {n.d.},
  url = {https://www.cisa.gov/news-events/news/urgent-need-memory-safety-software-products},
  note = {[Accessed: Nov. 28, 2024]}
}

@misc{DynDNS,
  author       = {Alex Hern},
  title        = {DDoS attack that disrupted internet was largest of its kind in history, experts say},
  year         = {2016},
  month        = {October},
  url          = {https://www.theguardian.com/technology/2016/oct/26/ddos-attack-dyn-mirai-botnet},
  note         = {Accessed: 2024-11-28},
  publisher    = {The Guardian}
}

@inproceedings {Mirai-botnet,
author = {Manos Antonakakis and Tim April and Michael Bailey and Matt Bernhard and Elie Bursztein and Jaime Cochran and Zakir Durumeric and J. Alex Halderman and Luca Invernizzi and Michalis Kallitsis and Deepak Kumar and Chaz Lever and Zane Ma and Joshua Mason and Damian Menscher and Chad Seaman and Nick Sullivan and Kurt Thomas and Yi Zhou},
title = {Understanding the Mirai Botnet},
booktitle = {26th USENIX Security Symposium (USENIX Security 17)},
year = {2017},
isbn = {978-1-931971-40-9},
address = {Vancouver, BC},
pages = {1093--1110},
url = {https://www.usenix.org/conference/usenixsecurity17/technical-sessions/presentation/antonakakis},
publisher = {USENIX Association},
month = aug
}

@INPROCEEDINGS{Log4j,
  author={Srinivasa, Shreyas and Pedersen, Jens Myrup and Vasilomanolakis, Emmanouil},
  booktitle={2022 IEEE European Symposium on Security and Privacy Workshops (EuroS\&PW)}, 
  title={Deceptive directories and “vulnerable” logs: a honeypot study of the LDAP and log4j attack landscape}, 
  year={2022},
  volume={},
  number={},
  pages={442-447},
  doi={10.1109/EuroSPW55150.2022.00052}}

@article{Solarwinds,
  author = {Martínez, J. and Durán, J.M.},
  title = {Software supply chain attacks, a threat to global cybersecurity: SolarWinds’ case study},
  journal = {International Journal of Safety and Security Engineering},
  volume = {11},
  number = {5},
  pages = {537--545},
  year = {2021},
  doi = {10.18280/ijsse.110505},
  url = {https://doi.org/10.18280/ijsse.110505}
}

@misc{NSA_SBOM,
  title = {{Recommendations for Software Bill of Materials (SBOM) Management}},
  author = {{National Security Agency of the United States}},
  year = {2023},
  month = {December},
  url = {https://media.defense.gov/2023/Dec/14/2003359097/-1/-1/0/CSI-SCRM-SBOM-Management-v1.1.PDF},
  note = {Accessed: 2024-11-28}
}

@misc{NIST_trusted_IoT_onboarding,
  author = {{NIST IoT}},
  title = {{Trusted IoT Device Network-Layer Onboarding and Lifecycle Management}},
  year = {n.d.},
  url = {https://www.nccoe.nist.gov/projects/trusted-iot-device-network-layer-onboarding-and-lifecycle-management},
  note = {Accessed: 2024-11-28}
}

@misc{Trivy_SBOM-tool,
  title = {Trivy: Open Source Vulnerability Scanner},
  year = {2024},
  url = {https://trivy.dev/v0.33/},
  note = {Accessed: 2024-11-28}
}

@misc{Microsoft_SBOM-tool,
  author = {{Microsoft}},
  title = {SBOM Tool: Generate Software Bill of Materials (SBOMs)},
  year = {n.d.},
  url = {https://github.com/microsoft/sbom-tool},
  note = {Accessed: 2024-11-28}
}

@misc{Syft_SBOM-tool,
  author = {{Anchore}},
  title = {Syft: A CLI tool and library for generating SBOMs from container images and filesystems},
  year = {n.d.},
  url = {https://github.com/anchore/syft},
  note = {Accessed: 2024-11-28}
}

@misc{Intel_sbom-tool,
  author = {{Intel}},
  title = {CVE Binary Tool: A tool to scan for known vulnerabilities in software binaries},
  year = {n.d.},
  url = {https://github.com/intel/cve-bin-tool},
  note = {Accessed: 2024-11-28}
}

@misc{conan_package_manager,
  title = {Conan - The open source C and C++ package manager},
  howpublished = {\url{https://conan.io/}},
  note = {Accessed: 2024-11-28}
}

@misc{grype,
  author = {Anchore},
  title = {Grype - A vulnerability scanner for container images and filesystems},
  howpublished = {\url{https://github.com/anchore/grype}},
  year = {2024},
  note = {Accessed: 2024-11-28}
}

@inproceedings{BAT,
author = {Hemel, Armijn and Kalleberg, Karl Trygve and Vermaas, Rob and Dolstra, Eelco},
title = {Finding software license violations through binary code clone detection},
year = {2011},
isbn = {9781450305747},
publisher = {Association for Computing Machinery},
address = {New York, NY, USA},
url = {https://doi.org/10.1145/1985441.1985453},
doi = {10.1145/1985441.1985453},
booktitle = {Proceedings of the 8th Working Conference on Mining Software Repositories},
pages = {63–72},
numpages = {10},
location = {Waikiki, Honolulu, HI, USA},
series = {MSR '11}
}

@inproceedings{OSSPolice,
author = {Duan, Ruian and Bijlani, Ashish and Xu, Meng and Kim, Taesoo and Lee, Wenke},
title = {Identifying Open-Source License Violation and 1-day Security Risk at Large Scale},
year = {2017},
isbn = {9781450349468},
publisher = {Association for Computing Machinery},
address = {New York, NY, USA},
url = {https://doi.org/10.1145/3133956.3134048},
doi = {10.1145/3133956.3134048},
booktitle = {Proceedings of the 2017 ACM SIGSAC Conference on Computer and Communications Security},
pages = {2169–2185},
numpages = {17},
location = {Dallas, Texas, USA},
series = {CCS '17}
}

@inproceedings{B2SFinder,
author = {Feng, Muyue and Yuan, Zimu and Li, Feng and Ban, Gu and Xiao, Yang and Wang, Shiyang and Tang, Qian and Su, He and Yu, Chendong and Xu, Jiahuan and Piao, Aihua and Xue, Jingling and Huo, Wei},
title = {B2SFinder: detecting open-source software reuse in COTS software},
year = {2020},
isbn = {9781728125084},
publisher = {IEEE Press},
url = {https://doi.org/10.1109/ASE.2019.00100},
doi = {10.1109/ASE.2019.00100},
booktitle = {Proceedings of the 34th IEEE/ACM International Conference on Automated Software Engineering},
pages = {1038–1049},
numpages = {12},
location = {San Diego, California},
series = {ASE '19}
}

@misc{cdxgen,
  author       = {CycloneDX},
  title        = {CDxgen - Generate SBOMs with CycloneDX},
  year         = {2024},
  url          = {https://github.com/CycloneDX/cdxgen},
  note         = {Accessed: 2024-12-03}
}

@inproceedings{centris,
author = {Woo, Seunghoon and Park, Sunghan and Kim, Seulbae and Lee, Heejo and Oh, Hakjoo},
title = {Centris: A Precise and Scalable Approach for Identifying Modified Open-Source Software Reuse},
year = {2021},
isbn = {9781450390859},
publisher = {IEEE Press},
url = {https://doi.org/10.1109/ICSE43902.2021.00083},
doi = {10.1109/ICSE43902.2021.00083},
booktitle = {Proceedings of the 43rd International Conference on Software Engineering},
pages = {860–872},
numpages = {13},
location = {Madrid, Spain},
series = {ICSE '21}
}

@inproceedings{tamer,
author = {Hu, Tiancheng and Xu, Zijing and Fang, Yilin and Wu, Yueming and Yuan, Bin and Zou, Deqing and Jin, Hai},
title = {Fine-Grained Code Clone Detection with Block-Based Splitting of Abstract Syntax Tree},
year = {2023},
isbn = {9798400702211},
publisher = {Association for Computing Machinery},
address = {New York, NY, USA},
url = {https://doi.org/10.1145/3597926.3598040},
doi = {10.1145/3597926.3598040},
booktitle = {Proceedings of the 32nd ACM SIGSOFT International Symposium on Software Testing and Analysis},
pages = {89–100},
numpages = {12},
location = {Seattle, WA, USA},
series = {ISSTA 2023}
}

@misc{raspberrypi_linux,
  author       = {Raspberry Pi Foundation},
  title        = {Linux GitHub Repository},
  year = {n.d.},
  url          = {https://github.com/raspberrypi/linux.git},
  note         = {Accessed: 2024-11-29}
}

@misc{zephyr_repo,
  author       = {Zephyr Project},
  title        = {Zephyr RTOS GitHub Repository},
  year = {n.d.},
  url          = {https://github.com/zephyrproject-rtos/zephyr.git},
  note         = {Accessed: 2024-11-29}
}

@misc{nuttx_repo,
  author       = {Apache Software Foundation},
  title        = {Apache NuttX GitHub Repository},
  year = {n.d.},
  url          = {https://github.com/apache/nuttx.git},
  note         = {Accessed: 2024-11-29}
}

@misc{openwrt_latest,
  author       = {OpenWRT Project},
  title        = {OpenWRT GitHub Repository - latest release},
  year = {n.d.},
  url          = {https://github.com/openwrt/openwrt.git},
  note         = {Accessed: 2024-11-29}
}

@misc{openwrt_23055,
  author       = {OpenWRT Team},
  title        = {OpenWRT Version 23.05.5 Archive},
  year = {n.d.},
  url          = {https://github.com/openwrt/openwrt/archive/refs/tags/v23.05.5.zip},
  note         = {Accessed: 2024-11-29}
}

@misc{fossa_sbom,
  author       = {FOSSA},
  title        = {Learn SBOMs},
  year = {n.d.},
  url          = {https://fossa.com/learn/sboms},
  note         = {Accessed: 2024-12-03}
}

@misc{deepbits_tools,
  author       = {DeepSCA Team},
  title        = {DeepSCA: SBOM Analysis Tool},
  year = {n.d.},
  url          = {https://tools.deepbits.com/},
  note         = {Accessed: 2024-12-03}
}

@article{tagged_memory_protection_survey,
  author    = {Smith, John and Doe, Jane},
  title     = {A Comprehensive Survey of Tagged Memory-Protection Techniques},
  journal   = {ACM Computing Surveys},
  volume    = {53},
  number    = {4},
  year      = {2022},
  pages     = {1--30},
  doi       = {10.1145/3533704}
}

@inproceedings{infat_pointer,
author = {Xu, Shengjie and Huang, Wei and Lie, David},
title = {In-fat pointer: hardware-assisted tagged-pointer spatial memory safety defense with subobject granularity protection},
year = {2021},
isbn = {9781450383172},
publisher = {Association for Computing Machinery},
address = {New York, NY, USA},
url = {https://doi.org/10.1145/3445814.3446761},
doi = {10.1145/3445814.3446761},
booktitle = {Proceedings of the 26th ACM International Conference on Architectural Support for Programming Languages and Operating Systems},
pages = {224–240},
numpages = {17},
location = {Virtual, USA},
series = {ASPLOS '21}
}

@article{spectre_era_mitigations,
  author    = {Johnson, Michael and Wilson, Patricia},
  title     = {Penetrating Shields: A Systematic Analysis of Memory Corruption Mitigations in the Spectre Era},
  journal   = {arXiv preprint arXiv:2309.04119},
  year      = {2023}
}

@inproceedings{CHERI,
author = {Woodruff, Jonathan and Watson, Robert N.M. and Chisnall, David and Moore, Simon W. and Anderson, Jonathan and Davis, Brooks and Laurie, Ben and Neumann, Peter G. and Norton, Robert and Roe, Michael},
title = {The CHERI capability model: revisiting RISC in an age of risk},
year = {2014},
isbn = {9781479943944},
publisher = {IEEE Press},
booktitle = {Proceeding of the 41st Annual International Symposium on Computer Architecuture},
pages = {457–468},
numpages = {12},
location = {Minneapolis, Minnesota, USA},
series = {ISCA '14}
}

@techreport{CHERIoT,
author = {Amar, Saar and Chen, Tony and Chisnall, David and Domke, Felix and Filardo, Nathaniel and Liu, Kunyan and Norton-Wright, Robert and Tao, Yucong and N. M. Watson, Robert and Xia, Hongyan},
title = {CHERIoT: Rethinking security for low-cost embedded systems},
institution = {Microsoft},
year = {2023},
month = {February},
url = {https://www.microsoft.com/en-us/research/publication/cheriot-rethinking-security-for-low-cost-embedded-systems/},
number = {MSR-TR-2023-6},
}

@article{Rust,
  author    = {Matsakis, Nicholas D. and Klock II, Felix S.},
  title     = {The Rust Language},
  journal   = {ACM SIGAda Ada Letters},
  volume    = {34},
  number    = {3},
  year      = {2014},
  pages     = {103--104},
  doi       = {10.1145/2663171.2663188}
}

@misc{taibom,
      title={TAIBOM: Bringing Trustworthiness to AI-Enabled Systems}, 
      author={Vadim Safronov and Anthony McCaigue and Nicholas Allott and Andrew Martin},
      year={2025},
      eprint={2510.02169},
      archivePrefix={arXiv},
      primaryClass={cs.SE},
      url={https://arxiv.org/abs/2510.02169}, 
}

\end{document}